\documentstyle[12pt]{article}
\begin{document}
\baselineskip 0.7cm

\newcommand{\gsim}{ \mathop{}_{\textstyle \sim}^{\textstyle >} }
\newcommand{\lsim}{ \mathop{}_{\textstyle \sim}^{\textstyle <} }
\newcommand{\vev}[1]{ \left\langle {#1} \right\rangle }
\newcommand{\lsp}{ \left ( }
\newcommand{\rsp}{ \right ) }
\newcommand{\lmp}{ \left \{ }
\newcommand{\rmp}{ \right \} }
\newcommand{\llp}{ \left [ }
\newcommand{\rlp}{ \right ] }
\newcommand{\labs}{ \left | }
\newcommand{\rabs}{ \right | }
\newcommand{\KEV}{ {\rm keV} }
\newcommand{\MEV}{ {\rm MeV} }
\newcommand{\GEV}{ {\rm GeV} }
\newcommand{\TEV}{ {\rm TeV} }
\newcommand{\mgut}{M_{GUT}}
\newcommand{\mint}{M_{I}}
\newcommand{\mgra}{M_{3/2}}
\newcommand{\mll}{m_{\tilde{l}L}^{2}}
\newcommand{\mdr}{m_{\tilde{d}R}^{2}}
\newcommand{\mllXX}[1]{m_{\tilde{l}L , {#1}}^{2}}
\newcommand{\mdrXX}[1]{m_{\tilde{d}R , {#1}}^{2}}
\newcommand{\mgy}{m_{G1}}
\newcommand{\mgl}{m_{G2}}
\newcommand{\mgc}{m_{G3}}
\newcommand{\nuR}{\nu_{R}}
\newcommand{\slL}{\tilde{l}_{L}}
\newcommand{\slLi}{\tilde{l}_{Li}}
\newcommand{\sdR}{\tilde{d}_{R}}
\newcommand{\sdRi}{\tilde{d}_{Ri}}

\begin{titlepage}

\begin{flushright}
TU-461
\\
June, 1994
\end{flushright}

\vskip 0.35cm
\begin{center}
{\large \bf Suppression of Proton Decay in the Missing-Partner Model for
Supersymmetric SU(5) GUT}
\vskip 1.2cm
J.~Hisano\footnote
{Fellow of the Japan Society for the Promotion of Science.}
, T.~Moroi$^1$
, K.~Tobe, and T.~Yanagida

\vskip 0.4cm

{\it Department of Physics, Tohoku University,\\
     Sendai 980-77, Japan}

\vskip 1.5cm

\abstract{
We construct a missing-partner model for supersymmetric SU(5) GUT
assuming the Peccei-Quinn symmetry, in which the SU(5) gauge coupling constant
remains in the perturbative regime below the gravitational scale
$\sim2.4\times 10^{18}\GEV$. The Peccei-Quinn symmetry suppresses the
dangerous dimension-five operators for the nucleon decay much below the
limit from the present proton-decay experiments. We also stress that due
to this suppression mechanism our model can accommodate even the large
$\tan\beta_H$ ($\sim$ 60) scenario which has been recently suggested to
explain the observed value of the $m_b/m_\tau$ ratio.
}

\end{center}
\end{titlepage}

%
%
%
%

Grand unified theories (GUT's) of strong and electroweak
interactions~\cite{PRL32-438} are based on the assumption of a large
hierarchy between two mass scales, $M_{GUT}\sim10^{16}\GEV$ and
$M_{\rm weak}\sim10^{2}\GEV$. In constructing a realistic GUT model, a serious
problem arises from a phenomenological requirement that the ${\rm
SU(2)}_L$-doublet Higgs $H_f$ must have a mass of order of the
electroweak scale to cause the breaking of ${\rm SU(2)}_L\times{\rm
U(1)}_Y$ symmetry, while the mass of the color-triplet Higgs $H_c$
should be at the GUT scale in order to guarantee the observed stability
of proton. This problem is not easily solved even in the supersymmetric
(SUSY) extension~\cite{NPB193-150} of GUT's, since an
extremely precise adjustment of parameters in the superpotential is required
to achieve such a large mass splitting of the doublet- and triplet-Higgs
multiplets.  Although such a tree-level hierarchy survives quantum
corrections, thanks to the cancellation in the SUSY
theories~\cite{ACTB12-437}, the required fine-tuning of parameters seems
very much unlikely.

The missing-partner model~\cite{PLB115-380} for the SUSY SU(5) GUT is a
well-known example in which the above doublet-triplet splitting is
naturally obtained without fine-tuning of parameters. However, this
model becomes strongly interacting below the gravitational scale
$M=M_{pl}/\sqrt{8\pi}\simeq2.4\times10^{18}\GEV$, since it contains somewhat
high-rank representations of SU(5) for Higgs multiplets, ${\bf 75,
50}$, and ${\bf \overline{50}}$.
Therefore, the perturbative description of GUT's is broken down before
reaching supergravity. A solution to this problem may be given if one
puts the mass of ${\bf 50+\overline{50}}$ at the gravitational scale $M$. In
this case, however, the color-triplet Higgses $H_{c}$ and $\overline{H}_{c}$
have a
relatively smaller mass $M_{H_{c}}\sim M_{GUT}^2/M \sim 10^{(14-15)}\GEV$
which has been already excluded~\cite{NPB402-46,nath} by the
present proton-decay experiments~\cite{PDG}.

In this letter we show that introduction of the Peccei-Quinn
symmetry~\cite{PRL38-1440} solves all problems mentioned above in the
missing-partner model.\footnote
{In the previous article~\cite{PLB291-263}, two of the authors (J.H.
and T.Y.)  consider the minimum SUSY-SU(5) model with the Peccei-Quinn
symmetry. As stressed later, there is a crucial phenomenological
difference between the previous and the present models.}
That is, the Peccei-Quinn symmetry suppresses the $H_{c}$-mediated
dimension-five ($D$=5) operators~\cite{NPB197-533} for the nucleon decay
much well below the present experimental limit even for the relatively
light color-triplet Higgses ($M_{H_{c}}\sim 10^{(14-15)}\GEV$). We stress that
this suppression also allows us to have a large $\tan\beta_H$ ($\sim$ 60)
which has been suggested as one of the parameter regions giving the
correct $m_b / m_\tau$ ratio~\cite{PRD49-1454}.

The original missing-partner model~\cite{PLB115-380} in the SUSY-SU(5)
GUT consists of the following chiral supermultiplets;
\begin{eqnarray}
\psi_i({\bf 10}),~~~\phi_i({\bf \overline{5}}),
{}~~~H({\bf 5}),~~~\overline{H}({\bf \overline{5}}),
{}~~~\theta({\bf 50}),~~~\overline{\theta}({\bf \overline{50}}),
{}~~~\Sigma({\bf 75}),
\label{contents}
\end{eqnarray}
where $i$(=1--3) represents family index. In this model we incorporate a
Peccei-Quinn symmetry ${\rm U(1)}_{PQ}$ under which the chiral
multiplets in Eq.~(\ref{contents}) transform as
\begin{eqnarray}
\psi_i({\bf 10}) &\rightarrow& e^{i\alpha/2} \psi_i({\bf 10}),
\nonumber \\
\phi_i({\bf \overline{5}}) &\rightarrow& e^{i\beta/2} \phi_i({\bf
\overline{5}}),
\nonumber \\
H({\bf 5}) &\rightarrow& e^{-i\alpha} H({\bf 5}),
\nonumber \\
\overline{H}({\bf \overline{5}}) &\rightarrow& e^{-i\frac{\alpha+\beta}{2}}
\overline{H}({\bf \overline{5}}),
\nonumber \\
\theta({\bf 50}) &\rightarrow& e^{i\alpha} \theta({\bf 50}),
\nonumber \\
\overline{\theta}({\bf \overline{50}}) &\rightarrow&
e^{i\frac{\alpha+\beta}{2}}
\overline{\theta}({\bf \overline{50}}),
\nonumber \\
\Sigma({\bf 75}) &\rightarrow& \Sigma({\bf 75}),
\label{PQ_charge}
\end{eqnarray}
with $3 \alpha+\beta\ne 0.$\footnote
{This is required for the Peccei-Quinn mechanism to work. On the other
hand, the baryon-number violating $D=5$
operators, $\phi_i \psi_j \psi_k \psi_l$, are forbidden under this condition,
irrespective of their origins.}
These ${\rm U(1)}_{PQ}$ charges are chosen such that the following
superpotential is allowed,
\begin{eqnarray}
\label{superpotential1}
W &=&
\frac{1}{4} h_{ij} \psi_i^{(AB)} \psi_j^{(CD)} H^{E}\epsilon_{ABCDE}
+ \sqrt{2} f_{ij} \psi_i^{(AB)} \phi_{Aj} \overline{H}_B
\nonumber \\ &&
+ G_H H^A \Sigma_{(FG)}^{(BC)} \theta^{(DE)(FG)} \epsilon_{ABCDE}
+ G_{\overline{H}} \overline{H}_A
\Sigma^{(FG)}_{(BC)}\overline{\theta}_{(DE)(FG)} \epsilon^{ABCDE}
\nonumber\\ &&
+M_{75}\Sigma_{(CD)}^{(AB)}\Sigma_{(AB)}^{(CD)}
-\frac13 \lambda_{75}
\Sigma_{(EF)}^{(AB)}\Sigma_{(AB)}^{(CD)}\Sigma_{(CD)}^{(EF)},
\end{eqnarray}
and that the Peccei-Quinn symmetry is not broken by $\vev{\Sigma}\neq
0$. Here, indices $A, B, C \dots$ are the SU(5) indices which run from
1 to 5,
$\epsilon_{ABCDE}$  and $\epsilon^{ABCDE}$ are the fifth-antisymmetric
tensors, and the indices in $(AB)$ are antisymmetric.

The above model is still incomplete, since $\theta({\bf 50})$ and
$\overline{\theta}({\bf \overline{50}})$ can not have an invariant mass term.
To
give large masses to them we double the $H$ and $\theta$ sector
introducing a new set of chiral multiplets, $\overline{H}^{'}({\bf
\overline{5}})$, $H^{'}({\bf 5})$,
$\overline{\theta}^{'}({\bf \overline{50}})$, and
$\theta^{'}({\bf 50})$ , which have opposite ${\rm U(1)}_{PQ}$
charges of the corresponding original fields, $H$, $\overline{H}$, $\theta$,
and
$\overline{\theta}$. We add a new superpotential to
Eq.~(\ref{superpotential1}),
\begin{eqnarray}
\label{superpotential2}
W^{'} &=&
 G_H^{'} H^{\prime A} \Sigma_{(FG)}^{(BC)} \theta^{\prime(DE)(FG)}
\epsilon_{ABCDE}
+G_{\overline{H}}^{\prime} \overline{H}_A^{\prime} \Sigma^{(FG)}_{(BC)}
\overline{\theta}_{(DE)(FG)}^{\prime} \epsilon^{ABCDE}
\nonumber \\ &&
+ M_1 \overline{\theta}_{(AB)(CD)} \theta^{\prime(AB)(CD)} + M_2
\overline{\theta}_{(AB)(CD)}^{'} \theta^{(AB)(CD)}.
\end{eqnarray}
To avoid that the SU(5) gauge coupling constant blows up below the
gravitational scale $M$, we assume
\begin{eqnarray}
M_1,~M_2 \gsim 10^{18}\GEV.
\end{eqnarray}
In this letter, we take $M_1=M_2=M$($\equiv 2.4\times10^{18}$GeV) for
simplicity. Then, we have four Higgses, $H$, $\overline{H}$, $H^{'}$,
$\overline{H}^{'}$, and one Higgs $\Sigma$ much below the
gravitational scale $M$.

The {\bf 75}-dimension Higgs $\Sigma$
has the following vacuum-expectation value that causes the breaking
SU(5) $\rightarrow$
SU(3)$_C$$\times$SU(2)$_L$$\times$U(1)$_Y$,
\begin{eqnarray}
\langle \Sigma \rangle_{(\gamma\delta)}^{(\alpha\beta)}
&=& \frac12 \left\{
\delta_\gamma^\alpha \delta_\delta^\beta -
\delta_\delta^\alpha \delta_\gamma^\beta
\right\} V_{\Sigma},
\nonumber\\
\langle \Sigma \rangle_{(cd)}^{(ab)}
&=& \frac32 \left\{
\delta_c^a \delta_d^b -
\delta_d^a \delta_c^b
\right\} V_{\Sigma},
\\
\langle \Sigma \rangle_{(b \beta)}^{(a \alpha)}
&=& -\frac12 \left\{
\delta_b^a \delta_\beta^\alpha
\right\} V_{\Sigma},
\nonumber
\end{eqnarray}
where
\begin{equation}
V_{\Sigma}=\frac32 \frac{M_{75}}{\lambda_{75}}
\end{equation}
obtained from the superpotential Eq.~(\ref{superpotential1}).  Here,
$\alpha, \beta\dots$ are the SU(3)$_C$ indices and $a, b\dots$ the
SU(2)$_L$ indices. This vacuum-expectation value generates masses for the
color-triplet Higgses as (after integrating out the heavy fields, $\theta$,
$\overline{\theta}^\prime$ and $\theta^\prime$, $\overline{\theta}$),
\begin{eqnarray}
  M_{H_c} H_c^\alpha \overline{H}_{c\alpha}^{\prime}
+ M_{\overline{H}_c} H_c^{\prime \alpha} \overline{H}_{c\alpha},
\end{eqnarray}
with
\begin{eqnarray}
\label{coloredmasses}
M_{H_c} \simeq 48 G_H G_{\overline{H}}^{'} \frac{V_{\Sigma}^2}{M},~~~
M_{\overline{H}_c} \simeq 48 G_{\overline{H}}G_H^{'} \frac{V_{\Sigma}^2}{M}.
\end{eqnarray}
The four ${\rm SU(2)}_L$-doublet Higgses, $H_f$, $\overline{H}_f$,
$H_f^{\prime}$, and $\overline{H}_f^{\prime}$, remain massless.

In order to break the Peccei-Quinn symmetry, we introduce a pair of
SU(5)-singlet chiral multiplets $P$ and $Q$ whose ${\rm U(1)}_{PQ}$ charges are
chosen
as $P\rightarrow e^{-i\frac12(3\alpha+\beta)}P$ and $Q\rightarrow
e^{i\frac32(3\alpha+\beta)}Q$ so that the following
superpotential is allowed~\cite{PLB291-418}
\begin{eqnarray}
\label{PQpotential}
W^{''} = \frac{f}{M} P^3 Q + g_P \overline{H}_A^{\prime} H^{\prime A} P.
\label{yukawa}
\end{eqnarray}
We have a very flat scalar potential for $P$ and $Q$ as
\begin{eqnarray}
V(P,Q) = \frac{f^2}{M^2} |P|^6 + \frac{f^2}{M^2} |3P^2Q|^2.
\end{eqnarray}
As pointed out in Ref.~\cite{PLB291-418}, the introduction of negative
soft-SUSY breaking mass\footnote
{This negative soft SUSY breaking mass may be induced by radiative
corrections from the $j_{ij} N_i N_j P$ interactions given in
Eq.~(\ref{majorana}). See
Ref.~\cite{PLB291-418} for details.}
$\sim - m^2$ for $P$
induces very naturally the Peccei-Quinn symmetry breaking\footnote
{To generate non-vanishing vacuum-expectation value for $Q$, we have
assumed the other soft SUSY-breaking term $\sim (m/M)P^3 Q$ (see
Ref.~\cite{PLB291-418} in detail). Here, $Q$ denotes the scalar
component of the chiral multiplet $Q$.}
at the intermediate scale,\footnote
{The breaking scale of the Peccei-Quinn symmetry is bounded by
astrophysics and cosmology \cite{kim} as
\begin{equation}
10^{10}{\rm GeV}
\le \langle P \rangle, \langle Q \rangle \le
10^{13}{\rm GeV}.
\end{equation}
}
\begin{eqnarray}
\vev{P} \simeq \vev{Q} \simeq \sqrt{\frac{M m}{f}} \sim 10^{11} \GEV,
\end{eqnarray}
provided $m \sim 1\TEV$ and $f\sim 1$.\footnote
{With this charge assignment for $P$ and $Q$, the Higgses $H_f$ and
$\overline{H}_f$ receive a mass only from the following
superpotential,
\begin{equation}
\frac{P^2 Q}{M^2} \overline{H} H.
\end{equation}
This gives a small invariant mass $\mu$ for $H_f$ and $\overline{H}_f$ as
$\mu \sim 1$GeV for $\langle P \rangle$$\simeq$$\langle Q
\rangle$$\sim10^{12}$GeV, which is not excluded for
$\tan\beta_H<\sqrt2$ \cite{decamp}. However,
if one takes $f\sim 10^{-4}$, then $\langle P \rangle$ and $\langle Q
\rangle$ are $\sim 10^{13}$GeV. In this case the invariant mass $\mu$
becomes $O$(1)TeV.}
This Peccei-Quinn symmetry breaking produces
an intermediate-scale mass for a pair of light SU(2)$_L$-doublet Higgses,
$H_f^{\prime}$
and $\overline{H}_f^{\prime}$,
\begin{eqnarray}
M_{H_f^{'}} = g_P {\vev P},
\end{eqnarray}
through the Yukawa interaction in Eq.~(\ref{yukawa}).

So far we have two
independent charges $\alpha$ and $\beta$ defined in Eq.~(\ref{PQ_charge})
and hence there are two global U(1)'s. To eliminate one of them, we
introduce right-handed neutrino multiplets $N_i$ ({\bf 1}) \cite{MOD-A1-541}.
In fact, with two possible Yukawa couplings
\begin{eqnarray}
\label{majorana}
W^{'''} = k_{ij} N_i \phi_j H + j_{ij} N_i  N_j P,
\end{eqnarray}
we have only one ${\rm U(1)}_{PQ}$ and the charge $\alpha$ is fixed as
$\alpha = 3\beta$. The Yukawa couplings $j_{ij}N_i N_j P$
in Eq.~(\ref{majorana}) induce
Majorana masses for the right-handed neutrino multiplets
$N_i$, with $\langle P\rangle \ne 0$. Interesting is that the Majorana
masses for the right-handed
neutrinos are expected to be $O(10^{11})$ GeV, which naturally induce
very small masses of neutrinos through the cerebrated see-saw
mechanism~\cite{seesaw1} in a range of the MSW
solution~\cite{nc9c-17} to the solar neutrino problem.

The color-triplet Higgses have an off-diagonal element
in their mass matrix as
\begin{eqnarray}
\lsp \overline{H}_c, \overline{H}_c^{\prime} \rsp
\lsp
\begin{array} {cc}
M_{\overline{H}_c} & 0 \\
g_P\vev{P}      &M_{H_c}
\end{array}
\rsp
\lsp
\begin{array} {c}
H_c^{\prime}\\
H_c
\end{array}
\rsp.
\end{eqnarray}
The baryon-number violating $D=5$ operators~\cite{NPB197-533} mediated by the
color-triplet
Higgses are given by in the present model (see Fig.~1 (a))
\begin{eqnarray}
\label{missingD=5}
\frac{g_P\vev{P}}{M_{H_c} M_{\overline{H}_c}} \frac{1}{2\sqrt{2}}
f_{ij} h_{kl}
\lsp \phi_{F i} \psi_j^{(FA)} \rsp
\lsp \psi_k^{(BC)} \psi_l^{(DE)} \rsp
 \epsilon_{ABCDE}.
\end{eqnarray}
Notice that those in the minimum SUSY-SU(5) GUT are given by (see Fig.~1 (b))
\begin{eqnarray}
\label{minimumD=5}
\frac{1}{M_{H_c}} \frac{1}{2\sqrt{2}}
f_{ij} h_{kl} \lsp \phi_{F i} \psi_j^{(F A)} \rsp \lsp \psi_k^{(BC)}
\psi_l^{(DE)} \rsp \epsilon_{ABCDE}.
\end{eqnarray}
Thus we easily see that the $D=5$ operators in the present model are
more suppressed by a factor $M_{H_f^{\prime}}/M_{H_c}$ compared with in the
minimum SUSY-SU(5) GUT.

We are now at the point to show a crucial difference between our model and
the previous Peccei-Quinn extension~\cite{PLB291-263} of the minimum
SUSY-SU(5) GUT. The mass
spectrum above U(1)$_{PQ}$ breaking scale contains four
SU(2)$_L$-doublets Higgses and hence the success of the gauge coupling
unification in the
minimum SUSY-GUT may be lost in general. Thus, we have non-trivial
constraints on the Higgs masses from the gauge coupling unification,
which are different from the minimum SUSY-SU(5) GUT. The crucial point
is that these constraints in the present model are much different from those
obtained in the previous model~\cite{PLB291-263}, since the
SU(3)$_C$$\times$SU(2)$_L$$\times$U(1)$_Y$ components of $\Sigma$({\bf
75}) have different masses for each other due to
the own vacuum-expectation value.  This mass splitting gives large
threshold corrections to the
SU(3)$_C$$\times$SU(2)$_L$$\times$U(1)$_Y$ gauge coupling constants.

The running of the three gauge coupling constants at the one-loop
level is given by the
following solutions to the renormalization group
equations~\cite{PRL70-709},
\begin{eqnarray}
\alpha_3^{-1} (m_Z) &=& \alpha_{5}^{-1} (\Lambda)
+ \frac{1}{2\pi} \Bigg\{
\left( -2 - \frac{2}{3} N_g \right) \ln \frac{m_{SUSY}}{m_Z}
  \nonumber \\
&&
+ (-9 + 2 N_g) \ln \frac{\Lambda}{m_Z}
-4 \ln \frac{\Lambda}{M_V}  \nonumber\\
&&
        + 9 \ln \frac{\Lambda}{M_\Sigma}
        +   \ln \frac{\Lambda}{0.8 M_\Sigma}
        + 10\ln \frac{\Lambda}{0.4 M_\Sigma}
	+ 3 \ln \frac{\Lambda}{0.2 M_\Sigma} \nonumber\\
& &
 	+ \ln \frac{\Lambda}{M_{H_c}}
	+ \ln \frac{\Lambda}{M_{\overline{H}_c}} \Bigg\},
\label{alpha3}
\\
\alpha_2^{-1} (m_Z) &=& \alpha_{5}^{-1} (\Lambda)
+ \frac{1}{2\pi} \Bigg\{
\left( -\frac{4}{3} - \frac{2}{3} N_g - \frac{5}{6} \right)\ln
\frac{m_{SUSY}}{m_Z}
  \nonumber \\
&&
+ (-6 + 2 N_g + 1) \ln \frac{\Lambda}{m_Z}
-6 \ln \frac{\Lambda}{M_V} \nonumber\\
&&
	+ 16 \ln \frac{\Lambda}{M_\Sigma}
	+ 6  \ln \frac{\Lambda}{0.4 M_\Sigma}\nonumber\\
&&
 	+ \ln \frac{\Lambda}{M_{H_f^\prime}}
		\Bigg\},
\label{alpha2}
\\
\alpha_1^{-1} (m_Z) &=& \alpha_{5}^{-1} (\Lambda)
+ \frac{1}{2\pi} \Bigg\{
\left( -\frac{2}{3} N_g - \frac{1}{2} \right)\ln \frac{m_{SUSY}}{m_Z}
 \nonumber \\
&&
+ \left(2 N_g + \frac{3}{5} \right) \ln \frac{\Lambda}{m_Z}
-10 \ln \frac{\Lambda}{M_V} \nonumber\\
&&
	+ 10 \ln \frac{\Lambda}{0.8 M_\Sigma}
	+ 10 \ln \frac{\Lambda}{0.4 M_\Sigma}\nonumber\\
&&
 	+ \frac25 \ln \frac{\Lambda}{M_{H_c}}
	+ \frac25 \ln \frac{\Lambda}{M_{\overline{H}_c}}
	+ \frac35 \ln \frac{\Lambda}{M_{{H}_f^\prime}} \Bigg\},
\label{alpha1}
\end{eqnarray}
where $\alpha_5 \equiv g^2_5 /4\pi$ is the SU(5) gauge coupling
constant, $M_V$ the heavy gauge boson mass ($M_V=2\sqrt{15}g_5
V_{\Sigma}$), and $\Lambda$ the renormalization point which is taken
$\Lambda\gg M_{GUT}$.
Here, we have assumed that all superparticles in the SUSY-standard
model have a SUSY-breaking common mass $m_{SUSY}$ for simplicity, and
the mass splitting of $\Sigma$({\bf 75}) has been included.
The each SU(3)$_C$$\times$SU(2)$_L$$\times$U(1)$_Y$ components of
$\Sigma$ have the following masses;
\begin{eqnarray}
({\rm SU(3)}_C\times{\rm SU(2)}_L\times{\rm U(1)}_Y)
&& {\rm mass} \nonumber\\
({\bf 8},{\bf 3}, { 0}) && M_{\Sigma} \nonumber\\
({\bf 3},{\bf 1}, {\frac53})~,
({\bf \overline{3}},{\bf 1},{ -\frac53}) && \frac45 M_{\Sigma} \nonumber\\
({\bf 6},{\bf 2},{\frac56})~,
({\bf \overline{6}},{\bf 2},{-\frac56}) && \frac25 M_{\Sigma} \nonumber\\
({\bf 1},{\bf 1} ,{0}) && \frac25 M_{\Sigma} \nonumber\\
({\bf 8},{\bf 1}, {0}) && \frac15 M_{\Sigma} \nonumber\\
({\bf 3},{\bf 2}, {-\frac56})~,
({\bf {\overline{3}}},{\bf 2}, {\frac56}) && 0~(\mbox{\rm
Nambu-Goldstone multiplets})
\end{eqnarray}
where $M_{\Sigma}=10\lambda_{75} V_{\Sigma} /3$.\footnote
{The Numbu-Goldstone multiplets are absorbed to the gauge multiplets forming
massive
vector multiplets $V$ at the GUT scale.}
  By eliminating
$\alpha^{-1}_{5}$ from Eqs.~(\ref{alpha3}-\ref{alpha1}), we obtain
simple relations \cite{NPB402-46,PLB291-263}:
\begin{eqnarray}
\label{MHC}
(3 \alpha_2^{-1} - 2 \alpha_3^{-1} - \alpha_1^{-1}) (m_Z)
	&=& \frac{1}{2\pi} \Bigg\{
		\frac{12}{5} \, \ln \frac{M_{H_c}M_{\overline{H}_c}}{M_{H_f^\prime}m_Z}
		- 2 \, \ln \frac{m_{SUSY}}{m_Z} \nonumber\\
&&	         - \frac{12}{5} \, \ln (1.7\times 10^4)
\Bigg\},\\
\label{MVMSIGMA}
(5 \alpha_1^{-1} - 3 \alpha_2^{-1} - 2 \alpha_3^{-1}) (m_Z)
	&=& \frac{1}{2\pi} \Bigg\{
		12 \, \ln \frac{M_V^2 M_\Sigma}{m_Z^3}
		+ 8 \ln \frac{m_{SUSY}}{m_Z}  \nonumber\\
&& 		 + 36 \, \ln (1.4)
\Bigg\}.
\end{eqnarray}
Notice that the last terms in Eqs.~(\ref{MHC},\ref{MVMSIGMA}) come
from the mass splitting of $\Sigma$({\bf 75}), which makes a crucial
difference between the previous and the present models.\footnote
{The mass splitting of $\Sigma$({\bf 24}) in the minimum SUSY-SU(5) GUT
and in its Peccei-Quinn extension does not produces these constant terms
as pointed out in Ref.~\cite{NPB402-46,PLB291-263}}

To perform a quantitative analysis, we use the two-loop renormalization
group equations between the weak and the GUT scales.  Instead of the
common mass $m_{SUSY}$ of superparticles we have used the mass spectrum
estimated from the minimum supergravity \cite{NPB402-46,nojiri} to calculate
the one-loop
threshold correction at the SUSY-breaking scale. Using the experimental
data $\alpha^{-1}(m_Z)=127.9\pm 0.2$, $\sin^2\theta_W(m_Z)=0.2326\pm
0.0008$, and $\alpha_3(m_Z)=0.118\pm 0.007$~\cite{prd44-817},
\footnote{If one uses the recent experimental data
$\alpha^{-1}(m_Z)=127.9\pm 0.2$,
$\sin^2\theta_W(m_Z)=0.2314\pm 0.0004$, and
$\alpha_3(m_Z)=0.118\pm 0.007$ \cite{INS}, one gets
\begin{eqnarray}
1.4 \times 10^{17}~\GEV \leq
&\frac{M_{H_c}M_{\overline{H}_c}}{M_{H_f^\prime}}&
\leq 5.5 \times 10^{20}~\GEV,\nonumber\\
8.4 \times 10^{15}~\GEV \leq
&(M_V^2 M_\Sigma)^{1/3}&
\leq 2.6 \times 10^{16}~\GEV.\nonumber
\end{eqnarray}
However, we use the old data in the text for a comparison with the
previous result.} we obtain
\begin{eqnarray}
\label{missingMHC}
3.7 \times 10^{17}~\GEV \leq
&\frac{M_{H_c}M_{\overline{H}_c}}{M_{H_f^\prime}}&
\leq 3.8 \times 10^{21}~\GEV,\\
\label{missingMGUT}
6.8 \times 10^{15}~\GEV \leq
&(M_V^2 M_\Sigma)^{1/3}&
\leq 2.4 \times 10^{16}~\GEV.
\end{eqnarray}
This should be compared with the previous result in
Ref.~\cite{PLB291-263,NPB402-46},
\begin{eqnarray}
\label{minimumMHC}
2.2 \times 10^{13}~\GEV \leq
&\frac{M_{H_c}M_{\overline{H}_c}}{M_{H_f^\prime}}&
\leq 2.3 \times 10^{17}~\GEV,\\
\label{minimumMGUT}
9.5 \times 10^{15}~\GEV \leq
&(M_V^2 M_\Sigma)^{1/3}
& \leq 3.3 \times 10^{16}~\GEV.
\end{eqnarray}
The main reason for the different results comes from the presence of the
constant terms in Eqs.~(\ref{MHC}, \ref{MVMSIGMA}) which originate from the
mass
splitting of $\Sigma({\bf 75})$.  Notice that
Eq.~(\ref{missingMHC}) suggests $M_{H_c}\sim M_{\overline{H}_c}\sim
10^{(13-16)}$GeV for $M_{H_f^\prime}
\simeq10^{10}$GeV. This is very much consistent with
Eq.~(\ref{coloredmasses}) with $V_\Sigma\simeq10^{(15-16)}$GeV and
$G_H G_H^\prime\sim1$.

The $D=5$ operator in the minimum SUSY-SU(5) model is proportional to
$1/M_{H_c}$ as shown in Eq.~(\ref{minimumD=5}). The detailed analysis
on the nucleon-decay experiments
gives the lower limit on the color-triplet Higgs mass as $M_{H_c}\ge 5
\times 10^{15}$GeV~\cite{NPB402-46} in the minimum model. On the
other hand in the present model one can easily estimate the nucleon
decay rate by replacing $1/M_{H_c}$ by
$M_{H_f^\prime}/M_{H_c}M_{\overline{H}_c}$ (see Eq.~(\ref{missingD=5})). This
mass ratio is nothing but
one in Eq.~(\ref{missingMHC}) derived from the requirement of the
gauge-coupling unification. We find that the constraint
Eq.~(\ref{missingMHC}) is much weaker\footnote
{Comparing the upper limits on $M_{H_c} M_{\overline{H}_c} /
M_{\overline{H}_f^\prime}$
in Eq.~(\ref{missingMHC}) and in Eq.~(\ref{minimumMHC}), we see that the
$D$=5 operators in our model can be suppressed by a factor $\sim 10^{-4}$
compared with in the previous model.}
than that in Eq.~(\ref{minimumMHC}) and hence this model is still
consistent with the lower limit on the nucleon lifetime even for the
case of large $\tan\beta_{H}\equiv \langle H_f \rangle/ \langle \overline{H}_f
\rangle$.\footnote
{When $\tan\beta_H$ is large, the $D=5$ operators for
the nucleon decay are proportional to $\tan\beta_H$ \cite{NPB402-46,nath}.}

We show the evolution of the
SU(3)$_C$$\times$SU(2)$_L$$\times$U(1)$_Y$ and SU(5) gauge coupling constants
in Fig.~2 taking
$M_{H_c}=M_{\overline{H}_c}=10^{15}$GeV and $M_{H_f^\prime}=10^{10}$GeV for a
demonstrational purpose. We see that the unification of three gauge
coupling constants occurs around $10^{16}$GeV and the SU(5) gauge
coupling constant stays in the perturbative regime below the
gravitational scale $M\simeq 2.4\times 10^{18}$GeV.

In Fig.~3 we show the constraint from the present nucleon-decay
experiments, taking the same parameters for $M_{H_c}$ and
$M_{H_f^\prime}$ as above. To demonstrate how safe our model is, we have chosen
even the large $\tan \beta_H = 60$ and the largest hadron matrix element
$\beta=0.03$GeV$^3$ (see Ref.~\cite{NPB402-46} for notations).  One can
see that the superparticle masses below 1TeV are still allowed in the
present model. Thus, we stress that the intriguing idea of the Yukawa
coupling unification $h_t(M_{GUT})=h_b(M_{GUT})=h_\tau(M_{GUT})$
\cite{yukawa-u} (which implies the large
$\tan\beta_H\simeq 50-60$) is consistent not only with the observed fermion
masses, $m_b=(4.2-4.4)$GeV~\cite{PRD49-1454} and
$m_t=(160-190)$GeV~\cite{top}, but also with the present lower limit on the
nucleon lifetime.

In this letter we have shown that the Peccei-Quinn extension of the
missing-partner model in
SUSY-SU(5) GUT is consistent with the observed stability of the proton,
even if the masses of the unwanted ${\bf 50+\overline{50}}$ are lifted
up to the gravitational scale so that  the SU(5) gauge coupling
constant remains small enough for the perturbative description of GUT's. We
believe that the present model is worth being
pursued, since it is only a known, perturbative SUSY-GUT model which is
phenomenologically consistent and naturally realizes the
doublet-triplet splitting of Higgs multiplets.

\newpage

%
%
\newcommand{\Journal}[4]{{\sl #1} {\bf #2} {(#3)} {#4}}
\newcommand{\PL}{\sl Phys. Lett.}
\newcommand{\PR}{\sl Phys. Rev.}
\newcommand{\PRL}{\sl Phys. Rev. Lett.}
\newcommand{\NP}{\sl Nucl. Phys.}
\newcommand{\ZP}{\sl Z. Phys.}
\newcommand{\PTP}{\sl Prog. Theor. Phys.}
\newcommand{\NC}{\sl Nuovo Cimento}

\newpage

\section*{Figure Captions}
Fig.~1\\
(a) The Feynman diagram of the baryon-number violating $D=5$
operators in the present model. (b) The corresponding Feynman diagram in
the minimum SUSY-SU(5) GUT.
\\~\\
Fig.~2\\ The flows of the running gauge coupling constants of
SU(3)$_C$$\times$SU(2)$_L$$\times$U(1)$_Y$ and SU(5). Here, $M_{H_c}$
and $M_{\overline{H}_c}$ are taken at $10^{15}$GeV, and $M_{H_f^\prime}$
at $10^{10}$GeV.  We use $\alpha^{-1}(m_Z)=127.9\pm0.2$,
$\sin^2\theta_W(m_Z)=0.2326\pm0.0008$, and $\alpha_3(m_Z)=0.118\pm0.007$
\cite{prd44-817} for the initial condition. We assume the
SUSY-breaking scale $\sim$1TeV.
\\~~\\
Fig.~3\\ The lower bound of the superparticle masses from the negative
search of the nucleon decay \cite{PDG}. The horizontal axis is the wino mass
and
the vertical line the sfermion mass.  Here, $M_{H_c}$ and
$M_{\overline{H}_c}$ are taken at $10^{15}$GeV, and $M_{H_f^\prime}$ at
$10^{10}$GeV.  We take $\tan\beta_H$=60 for the ratio of the
vacuum-expectation values, $\langle H_f\rangle/ \langle \overline{H}_f
\rangle$, and $\beta=0.03$GeV$^3$ for the hadron matrix element. We also
show the lower bound in the minimum SUSY-SU(5) GUT by the dashed line,
taking 10$^{17}$GeV for $M_{H_c}$ and the same values for the other
parameters.  The dotted line is the lower bound of the chargino mass from
LEP \cite{LEP} and the dash-dotted line that of the squark mass from CDF
\cite{CDF}.

\end{document}